\documentclass[uspaper, 10pt, conference]{ieeeconf}      

\IEEEoverridecommandlockouts                              
\usepackage{graphics} 
\usepackage{epsfig} 
\usepackage{mathptmx} 
\usepackage{times} 
\usepackage{amsmath} 
\usepackage{amssymb}  

\newtheorem{defn}{Definition}
\newtheorem{thm}{Theorem}

\newtheorem{lem}{Lemma}

\renewcommand{\QED}{\QEDopen}

\title{\LARGE \bf
On computation of the total set of robust discrete-time PID
controllers}

\author{Naim~Bajcinca
\thanks{N. Bajcinca is with the German Aerospace Center (DLR), Institute of Robotics and Mechatronics in Oberpfaffenhofen,
Germany}}

\begin{document}

\maketitle
\thispagestyle{empty}
\pagestyle{empty}

\begin{abstract}
The problem of finding the set of all multi-model robust PID and
three-term stabilizers for discrete-time systems is solved in this
paper. The method uses the fact that decoupling of parameter space
at singular frequencies is invariant under a linear
transformation. The resulting stable regions are composed by
convex polygonal slices. The design problem includes the assertion
of intervals with stable polygons and the detection of stable
polygons. This paper completes the solutions to both problems.
\end{abstract}

\section{INTRODUCTION}
\label{sec:Introduction} It has been shown that the stabilizing
region for time-continuous PID-controllers is defined by a set of
convex polygonal slices normal to $k_P$ axis in the
$(k_P,~k_I,~k_D)$ parameter space. Different approaches which
prove this result use generalization of the Hermite-Biehler
theorem, see \cite{bhatta:book2000}, calculation of the real-axis
intersections of the Nyquist plot, see \cite{munro:2000} and
singular frequencies, see \cite{kaes:ecc2001},
\cite{bajc:auto:2004} and \cite{bajc:ifac2002}. In the cited
articles, it has been shown that the design of PID parameters due
to decoupling of PID controller parameters at singular frequencies
may be divided into two subproblems: (A) assertion of stable
intervals of parameter $k_P$ independently on parameters $k_I$ and
$k_D$ and (B) detection of stable polygons on the plane
($k_I,~k_D$) for a given $k_P$. the solutions to problems (A) and
(B) for the continuous-time systems can be found in
\cite{bajc:auto:2004} and \cite{bajc:med2001}.

In the present paper the time-continuous theory is transferred to
the design of PID and three-term controllers in discrete-time
domain. First related results are reported in
\cite{bajc:ifac2002}, where the set of all Schur-stable
controllers is shown to be composed by transformed polygonal
slices in a parameter space ($r_1,r_2,r_3$). While a solution to
the problem (B) as proposed in \cite{bajc:med2001} applies
directly in the discrete-time domain, see
Section~\ref{sec:stablePolygons}, this paper addresses the problem
(A) in Section~\ref{sec:StableGridding}.
Section~\ref{sec:basicDefinitions} proves that parameter
decoupling applies in general at singular frequencies. Basically,
a function is constructed from the characteristic polynomial, such
that its imaginary part on each point on unity circle depends only
on the gridding parameter (here conventionally chosen) $r_3$ and
is independent on parameters $r_1$ and $r_2$. This result will
lead to the conclusion that for a given plant a fixed number of
singular frequencies must be available for its stabilization.
Given that parameter $r_3$ defines uniquely the number of singular
frequencies, one can directly discriminate the intervals of
parameter $r_3$, which for no triple $(r_1,~r_2,~r_3)$ can provide
stabilization.  The simplicity of the criterion is, however,
achieved at the price of conservativeness, since just necessity is
provided. To cope with conservativeness the approach proposed in
\cite{bajc:auto:2004} for the continuous-time case may be directly
used. All aspects of the method are illustrated by examples.

\section{Problem definition}
\label{sec:problemDefinition}
Consider a simple closed curve $\Gamma~=\{z\mid
z=\tau(\alpha)+j~\eta(\alpha),~\alpha\in [a,~b]\}$, in $z-$plane,
which is symmetric to the real axis $\tau$ and can be expressed in
the form
\begin{equation}
    \label{eq:gammaeq}
    \Gamma:~~F(\tau, \eta)=0,
\end{equation}
and a characteristic equation of the form
\begin{equation}
\label{eq:basicEquation} p=A(z)Q(z,r_1,r_2,r_3)+B(z)=0,
\end{equation}
where $A$, $B$ are polynomials in $z$ and $r_1,~r_2,~r_3$ are real
parameters, that enter linearly in $Q$,
\begin{equation}
\label{eq:QEquation}
Q=\delta_1(z)r_1+\delta_2(z)r_2+\delta_3(z)r_3,
\end{equation}
where the polynomials $\delta_i(z)$ are assumed to be of the order
$\leq 2$ in $z$.

The following problem is handled in this article: Compute the set
of all parameters $r_1,r_2,r_3$, such that the polynomial
(\ref{eq:basicEquation}) is $\Gamma-$stable, that is, all its
eigenvalues are enclosed by $\Gamma$. Of main interest in this
article are circles with center on the real axis $\tau$ and an
arbitrary radius, which will be referred to as $\Gamma-$circles.
Especially important is the unity circle because of
Schur-stability. $\Gamma-$region refers to the region enclosed by
$\Gamma$.

It may be easily shown that (\ref{eq:basicEquation}) describes the
characteristic equation of a feedback loop with a PID or a
three-term controller. Indeed a discrete-time equivalent of the
PID controller has the transfer function
\begin{equation}\label{eq:PID}
C_1(z)=\frac{c_1 + c_2z + c_3z^2}{(z + z_1)(z - 1)}.
\end{equation}
\noindent Its structure follows in the quasi-continuous
consideration by applying the rectangular integration rule $(s
\rightarrow (z -1)/Tz)$ to the ideal PID controller $k_I/s + k_P +
k_Ds$, resulting in $z_1 = 0$, or by the trapezoidal integration
rule $(s \rightarrow 2(z - 1)/T(z + 1))$, resulting in $z_1 = 1$.
Also the realizable PID controller $k_I/s + k_P + k_Ds/(1 + T_1s)$
converts by the trapezoidal integration rule to the controller
structure (\ref{eq:PID}) with a pole at $z_1 = -(2T_1 - T)/(2T_1 +
T)$. The following derivation also holds for a three-term
controller with an arbitrary second order denominator polynomial
\begin{equation}\label{eq:threeTerm1}
 C_2(z)=\frac{n(z)(c_1 + c_2z + c_3z^2)}{d(z)}.
\end{equation}
For both controller structures, (\ref{eq:PID}) and
(\ref{eq:threeTerm1}), the polynomial $Q$ is of the form
\begin{equation}
\label{eq:SchurQ'Equationc1c2c3} Q=c_1+c_2 z +c_3 z^2.
\end{equation}
It is clear that (\ref{eq:QEquation}) and
(\ref{eq:SchurQ'Equationc1c2c3}) are connected via a parameter
space transformation
\begin{equation}\label{eq:Tmatrix}
c=T~r,~~\textrm{det~}T~\neq 0,
\end{equation}
with $c=[c_1,c_2,c_3]^T,~~r=[r_1,r_2,r_3]^T$. The rotation matrix
$T$ is determined by the polynomials,
$\delta_1(z),\delta_2(z),\delta_3(z)$.

\section{Basic definitions and theorems}
\label{sec:basicDefinitions}
Let $H$ and $G$ represent the real and imaginary part of the
characteristic  polynomial $p(z,r_1,r_2,r_3)$ in
(\ref{eq:basicEquation}) on $\Gamma$.

{\it
\begin{defn}\label{def:SingulaeresGammaGebiet} $\Gamma$ is
said to be singular with respect to parameters $r_1$ and $r_2$ in
(\ref{eq:basicEquation}) if
\begin{equation}\label{eq:RangebdingungFuerSingulaereGammaGebiete}
   \textrm{Rank}\frac{\partial{}(H,G)}{\partial{}(r_1,r_2)}=1\qquad \textrm{for
  all } z\in{}\Gamma{}.
 \end{equation}
\end{defn}
} The latter equation is referred to as the rank-condition. Unless
otherwise stated, $\Gamma$ will be assumed singular for the rest
of the article. A zero of the polynomial (\ref{eq:basicEquation}),
which additionally fulfills the rank-condition is referred to as
singular frequency. For a fixed parameter $r_3$, singular
frequencies are mapped to straight lines on the plane ($r_1, r_2$)
and the eigenvalues of (\ref{eq:basicEquation}) can cross over a
singular $\Gamma$ just at singular frequencies.

{\it
\begin{defn}\label{def:decouplingFunction}
A function $E_{\Gamma}(z)$ defined as
\begin{equation}\label{eq:decouplingFunction}
Q(z,r_1,r_2,r_3)=E_{\Gamma}(z)~q(z,r_1,r_2,r_3)
\end{equation}
with
\begin{equation}\label{eq:Iq1}
I_q=r_3 g_3(\alpha)+g_0(\alpha),~~~\alpha\in [a,~b]
\end{equation}
i.e. $\frac{\partial I_q}{\partial r_1}=\frac{\partial
I_q}{\partial r_2}=0$, where $I_q$ stands for the imaginary part
of $q$, will be referred to as decoupling function of $Q$ over
$\Gamma$.
\end{defn}
}

{\it
\begin{lem}\label{lem:derr1=derr2=0}
If $\Gamma$ is singular then
\begin{equation}\label{eq:derr1r2}
\frac{\partial I_q}{\partial r_1}=0 \Leftrightarrow \frac{\partial
I_q}{\partial r_2}=0,~~~\forall z\in \Gamma.
\end{equation}
\end{lem}
}

{\emph{Proof.~}} Consider the rank-condition
(\ref{eq:RangebdingungFuerSingulaereGammaGebiete}). It may be
checked that
\begin{displaymath}
\frac{\partial{}(H,G)}{\partial{}(r_1,r_2)}=\left(R^2_A+I^2_A\right)\left(R^2_\phi+I^2_\phi\right)\left(\frac{\partial
R_q}{\partial r_1} \frac{\partial I_q}{\partial
r_2}-\frac{\partial I_q}{\partial r_1}\frac{\partial R_q}{\partial
r_2}\right)
\end{displaymath}
If rank-condition applies, that is, the right-hand side of the
above equation is zero, then $\frac{\partial I_q}{\partial r_1}=0
\Rightarrow \frac{\partial I_q}{\partial r_2}=0$, since
$\frac{\partial R_q}{\partial r_1}\neq 0$. Similarly,
$\frac{\partial I_q}{\partial r_2}=0 \Rightarrow \frac{\partial
I_q}{\partial r_1}=0$.\hfill \QED

{\it
\begin{thm}\label{thm:decoupleFunction} Two decoupling functions of $Q$ in (\ref{eq:QEquation}) over
$\Gamma$ are simply given by
\begin{equation}\label{eq:decouplingFuntion}
E_\Gamma(z) = \delta_1(z),~~~E_\Gamma(z) = \delta_2(z).
\end{equation}
\end{thm}
}

{\emph{Proof.~}}Suppose $E_\Gamma(z)=\delta_1(z)$. Then
\begin{displaymath}
q = r_1+\frac{\delta_2(z)}{\delta_1(z)}r_2 +
\frac{\delta_3(z)}{\delta_1(z)}r_3,
\end{displaymath}
and $\frac{\partial I_q}{\partial r_1}\equiv 0$.
Lema~\ref{lem:derr1=derr2=0} guarantees that for all $z\in\Gamma$
the imaginary part of $\frac{\delta_2(z)}{\delta_1(z)}$ is also
zero, that is, (\ref{eq:Iq1}) applies. Since
$\frac{\delta_2(z)}{\delta_1(z)}$ is real on $\Gamma$, its inverse
will be also real, that is, the imaginary part of
$\frac{\delta_1(z)}{\delta_2(z)}$ for all $z\in\Gamma$ is zero.
Hence, $\delta_2(z)$ represents also a decoupling function.
\hfill\QED

The next statement may be directly checked.
 {\it
\begin{thm}\label{thm:decoupling}Consider the function
\begin{equation}\label{eq:Fz}
F(z):=\frac{p(z)}{A(z)E_\Gamma(z)}.
\end{equation}
The equation $F(z)=0$ for $z\in\Gamma$ decouples the parameters
$r_1, r_2$ and  $r_3$ into two equations
\begin{eqnarray}\label{eq:singFreqsLines}
r_1 h_1(\alpha)+r_2 h_2(\alpha)+h_0(\alpha)&=&0,\\
\label{eq:singFreqs}r_3 g_1(\alpha)+g_0(\alpha)&=&0.
\end{eqnarray}
\end{thm}
}

The  latter theorem is essential for solving the problem (A) as
defined in the introduction of the article. The first equation
refers to the generator of singular lines, and the second one to
the generator of singular frequencies.

Finally, let $\{p_{\nu}\}$ represent a finite set of polynomials
of the form (\ref{eq:basicEquation}), i.e.
\begin{eqnarray}
    \label{eq:basicEquationFamily}
p_{\nu}=A_{\nu}(z)Q(z,r_1,r_2,r_3)+B_{\nu}(z).
\end{eqnarray}
For example, this may be a multi-model of a continuum of plants or
Kharitonov polynomials of an interval uncertainty.

{\it
\begin{thm}The rank-condition (\ref{eq:RangebdingungFuerSingulaereGammaGebiete}) for the set
of polynomials (\ref{eq:basicEquationFamily}) does not depend on
the polynomials $A_{\nu}(z)$ and $B_{\nu}(z)$.
\end{thm}
}

Hence a singular $\Gamma$ is completely determined by the
polynomial $Q$ in (\ref{eq:QEquation}). Though the singularity of
$\Gamma$ is not destroyed by polynomials $A_{\nu}(z)$ and
$B_{\nu}(z)$, they influence the root-condition, i.e. the singular
frequencies on $\Gamma$.

\subsection{Schur-stability}
Consider the Schur-circle defined as
$\Gamma_1~=\{e^{j\alpha}:~\alpha\in [-\pi,~\pi]\}$. It can be
easily checked that the rank-condition
(\ref{eq:RangebdingungFuerSingulaereGammaGebiete}) is satisfied
over $\Gamma_1$ for
\begin{equation}
\label{eq:SchurQEquation} Q=(1+z^2) r_1+z r_2+ r_3
\end{equation}
and that the matrix $T$, as defined in (\ref{eq:Tmatrix}) is
\begin{eqnarray}\label{eq:Tcircle} {T} = \left[
\begin{array}{ccc}
1 & 0 & 1\\0 & 1 & 0\\1 & 0 & 0
\end{array}
\right].
\end{eqnarray}

According to Theorem~\ref{thm:decoupleFunction} the decoupling
functions of (\ref{eq:SchurQEquation}) over the Schur-circle
$\Gamma_1$ are
\begin{equation}\label{eq:decouplingFunctionScur}
E_\Gamma(z) = 1+z^2~~\textrm{or}~~E_\Gamma(z) = z.
\end{equation}
Let $E_\Gamma(z) = z$. It may be easily shown that the imaginary
part of the function
\begin{equation}\label{eq:FzSchur}
F(z)=\frac{1+z^2}{z} r_1+r_2 + \frac{B(z)}{z A(z)}r_3
\end{equation}
on the Schur-circle is of the form (\ref{eq:Iq1}), that is
imaginary part of $\frac{1+z^2}{z}$ is zero on the Schur-circle.

\subsection{$\Gamma$-stability}
\begin{figure}[hbt]
\begin{center}
\includegraphics[scale=0.45]{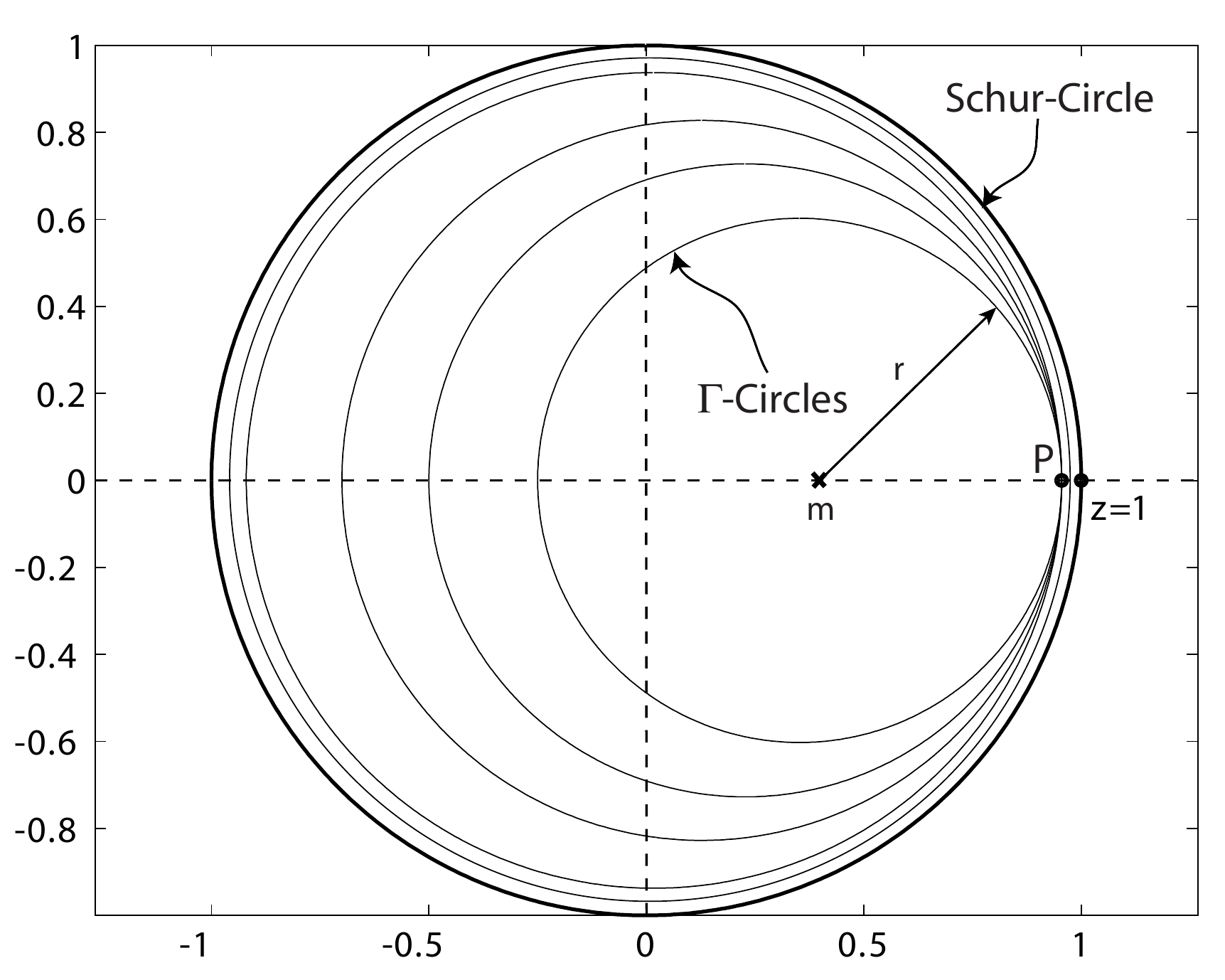}
\caption{Schur- and $\Gamma-$circles}\label{fig:circles}
\end{center}
\end{figure}

Consider a $\Gamma-$circle with center on real axis $\Gamma~=\{m+r
e^{j\alpha}$, $\alpha\in [-\pi,~\pi]\}$, Fig~\ref{fig:circles}.
Now it can be shown that
\begin{equation}
\label{eq:CirclesQEquation} Q=(r^2-m^2+z^2) r_1+(z-m) r_2+ r_3.
\end{equation}
For $\Gamma-$circles with center at $\tau=m$ and radius $r$ a
transformation matrix from $c-$ to $r-$parameter space is checked
to be,
\begin{eqnarray}\label{eq:12} {T} = \left[
\begin{array}{ccc}
r^2 - m^2 & -m & 1\\0 & 1 & 0\\1 & 0 & 0
\end{array}
\right]
\end{eqnarray}

A corresponding decoupling function is
\begin{equation}\label{eq:decouplingFunctionCircle}
E_\Gamma(z) = z-m
\end{equation}
and
\begin{equation}\label{eq:FzCircle}
F(z)=\frac{r^2-m^2+z^2}{z-m} r_1+r_2 + \frac{B(z)}{(z-m) A(z)}r_3.
\end{equation}

\subsection{Example}
{{Consider the discrete-time model of the plant
\begin{equation}\label{eq:GsDigit} G=10^{-6}\frac{4.165 z^3 +
45.77 z^2 + 45.77 z + 4.165}{z^4 - 3.985z^3 + 5.97z^2 - 3.985z +
1},
\end{equation}
and a three-term stabilizer
\begin{equation}
\label{eq:threeTerm} C(z)=10^4\frac{(z^2 - 1.541 z
+0.5992)(c_1+c_2 z+c_3 z^2)}{z(z+0.4047)(z+0.2162)(z-0.4934)},
\end{equation}
whose parameters $c_1, c_2, c_3$ should be synthesized. The
synthesis is done in $(r_1,r_2,r_3)$-parameter space. Therefore
the transformation (\ref{eq:Tcircle}) can be used. Then
\begin{eqnarray}\label{eq:Azexmaple}
A&=& {z}^{5}+9.44z^4-5.34 z^3-9.34z^2+5.04 z+0.59\\
\label{eq:Bzexmaple}
B&=&0.19 z^8-0.73 z^7+z^6-0.45 z^5-0.12 z^4+\cdots\nonumber\\
{} &{} & 0.14 z^3-0.009z^2-0.008z.
\end{eqnarray}

The equation (\ref{eq:singFreqs}) reads
\begin{equation}\label{eq:r3alpha}
r_3=\frac{C^{5}_\alpha- 2.71\, C^{4} _\alpha+ 1.61\, C^{3}_\alpha
+ 1.35\, C^{2} _\alpha- 1.70\,C _\alpha +
 0.44}{ 0.02\, C^{4} _\alpha+ 0.08
\, C^{3}_\alpha- 0.13\, C^{2} _\alpha - 0.08\,C_\alpha + 0.11}.
\end{equation}
where $C_\alpha=\cos(\alpha)$. For illustration purposes its plot
is shown in Fig.~\ref{fig:kpwexample}. E.g. for $r_3=-0.26118$,
the singular frequencies lying on the Schur-circle are computed to
be
\begin{eqnarray}
\begin{array}{lcl}\label{eq:sfs1}
\alpha'_1=0 &~~~\Rightarrow~~~&z'_1=1,\\
\alpha'_2=\pm0.4097&~~~\Rightarrow~~~&z'_{2}=0.9172\pm j 0.3983,\\
\alpha'_3=\pm 0.97300&~~~\Rightarrow~~~&z'_{3}=0.5628\pm j
0.8266,\\
\alpha'_4=\pm\pi&~~~\Rightarrow~~~&z'_{4}=-1.
\end{array}
\end{eqnarray}

\begin{figure}[h]
\begin{center}
{\includegraphics[scale=0.4]{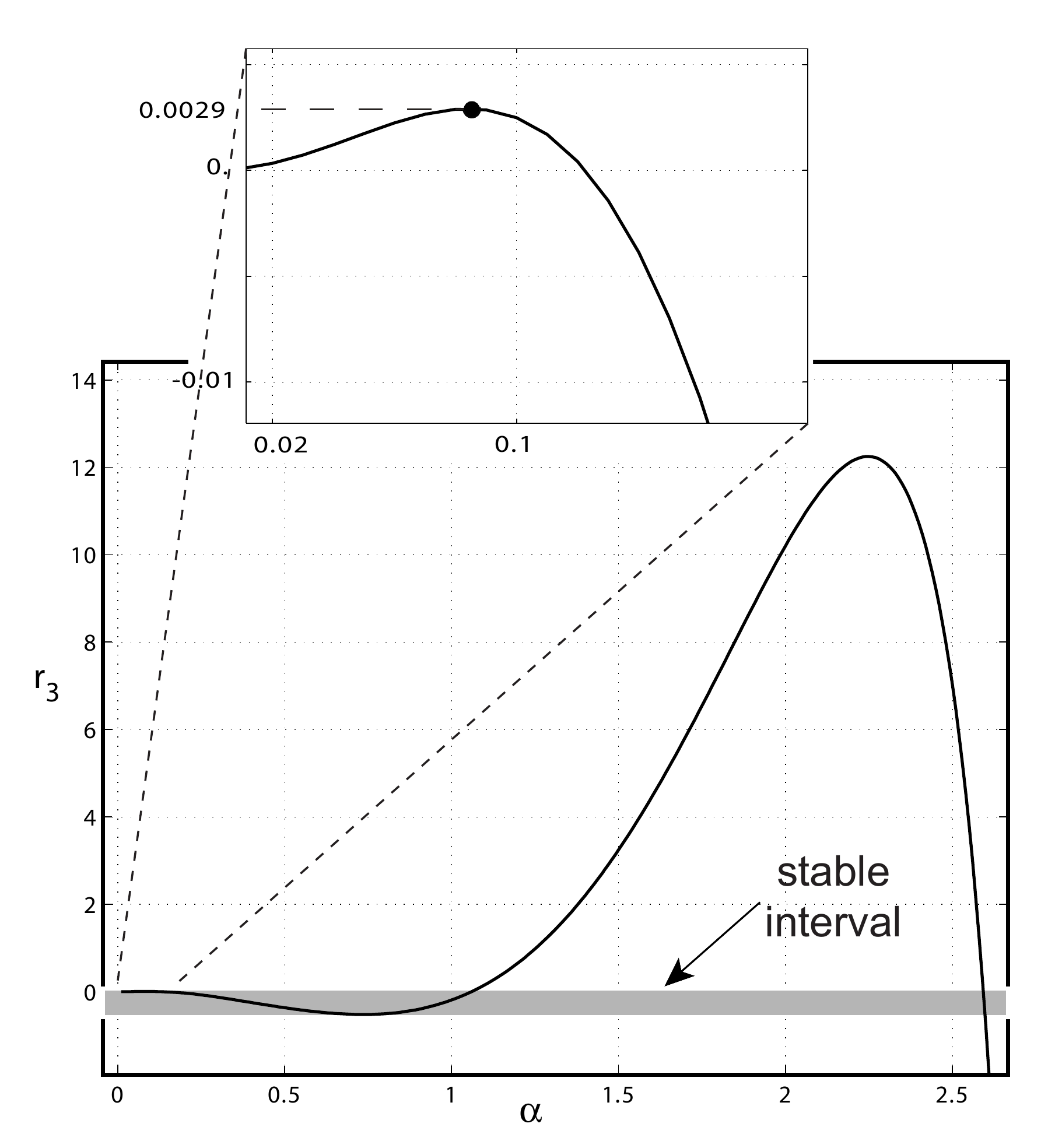}} \caption{The
generator fo singular frequencies; Equation~\ref{eq:r3alpha}}
\label{fig:kpwexample}
\end{center}
\end{figure}

}}

\section{Stable convex polygons}
\label{sec:stablePolygons}
Notice that the characteristic equation in
(\ref{eq:basicEquation}) over singular $\Gamma$ decouples
equivalently to the two equations in (\ref{eq:singFreqsLines}) and
(\ref{eq:singFreqs}). For a fixed $r_3$ the equation
(\ref{eq:singFreqsLines}) defines a set of singular frequencies
$\{\alpha'_{\nu}\}$ on $\Gamma$. The equation (\ref{eq:singFreqs})
determines the set of straight lines $\{\lambda_{\nu}\}$, which
are determined by $\{\alpha'_{\nu}\}$. Geometrically, the straight
lines $\{\lambda_{\nu}\}$ represent the eigenvalue boundaries in
the $(r_1,r_2)$-plane for the fixed parameter $r_3$. Thus, stable
regions are composed by convex polygons and the design problem
defined in the previous section is decoupled into two subproblems:
(A) assertion of stable intervals of parameter $r_3$ independently
on parameters $r_1$ and $r_2$ and (B) detection of stable polygons
on the plane ($r_1,~r_2$) for a given $r_3$.

For the sake of completion, this section recalls briefly the
solution of problem (B), while problem (A) will be thoroughly
discussed in the next section. For details the reader is referred
to \cite{bajc:med2001}, \cite{bajc:auto:2004}. The algorithm is
motivated by the concept of inner polygons, which claim a
necessary condition for stability: A polygon $\Pi$ is said to be
an inner polygon if any transition over $\lambda_\nu$ inside the
polygon causes an eigenvalue-pair to enter the $\Gamma-$region at
the corresponding frequency.

\begin{figure}[h]
\begin{center}
{\includegraphics[scale=0.7]{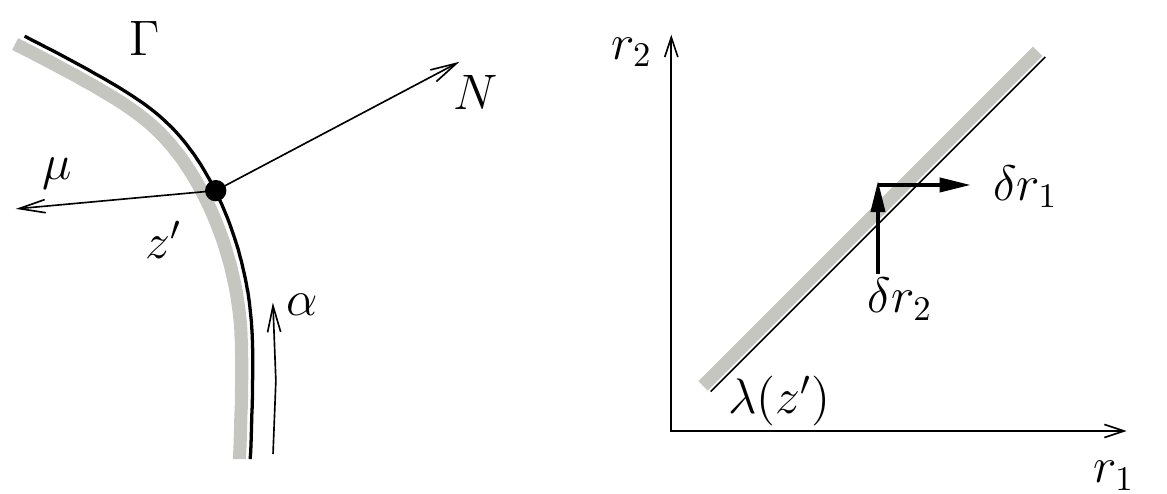}} \caption{Definition
of $e_1$ and $e_2$: the motion of eigenvalues in the vicinity of a
singular frequency $z'$. By convention, the shaded part refers to
the stable side of $\Gamma$ and the normal $N$ points outside the
$\Gamma$-region.} \label{fig:e1e2defs}
\end{center}
\end{figure}

In order to automate the detection of an inner polygon each
singular line $\lambda{}(\alpha')$, will be assigned a
"transition" function $e$: it is negative if the transition
$[\delta{}r_1,\delta{}r_2]$ (see Fig.~\ref{fig:e1e2defs}) over the
singular line causes an eigenvalue to become stable, otherwise it
is positive. Let $e_1$ correspond to $\delta{}r_1>0,~
\delta{}r_2=0$ and $e_2$ to $\delta{}r_2>0,~ \delta{}r_1=0$. In
\cite{bajc:med2001} it shown that the motion of eigenvalues at
$z'=\tau(\alpha')\pm j\eta(\alpha')$ defined as the scalar product
$e=\mu N$ (see Fig.~\ref{fig:e1e2defs}) is described by the
equation
\begin{equation}
    \label{eq:muexps}
    e_{1/2}:=\left(\frac{\partial
    F}{\partial\tau}\Bigg\vert\frac{\partial(H,G)}{\partial(\eta,r_{1/2})}
    \Bigg\vert+\frac{\partial F}{\partial\eta}\Bigg\vert\frac{\partial(H,G)}
    {\partial(r_{1/2},\tau)}\Bigg\vert\right)_{\normalsize
    \alpha'}.
\end{equation}
Using this information, an algorithm for the detection of the
inner polygons with maximal $\Gamma$-stable eigenvalues may be
developed, see \cite{bajc:med2001}. Such polygons are the only
candidates, which need to be checked for stability. To this end,
it suffices to check any point within such a polygon.

\begin{figure}[h]
\begin{center}
{\includegraphics[scale=0.45]{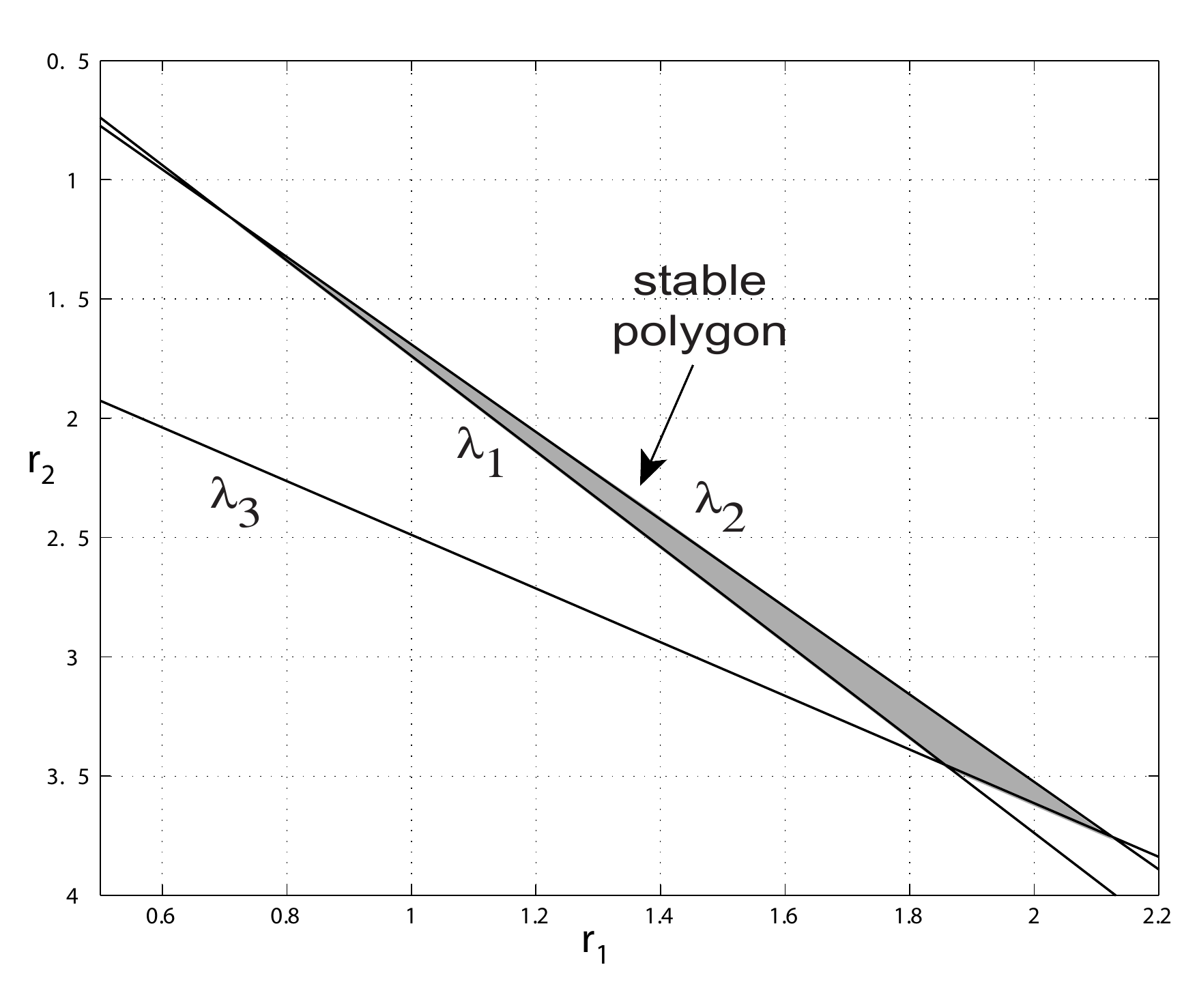}}
\caption{The stable polygon lying on the plane $r_3=-0.26118$}
\label{fig:astabpolyg1}
\end{center}
\end{figure}

\subsection{Example: (cont.)}
{~Reconsider the plane $r_3=-0.26118$. The corresponding singular
frequencies were computed in the previous section,
(\ref{eq:sfs1}). Each of these singular frequencies $z'_\nu$
generates a straight line $\lambda_\nu$ on the plane
$r_3=-0.26118$. The resulting stable polygon is shown in
Fig.~\ref{fig:astabpolyg1}. It is enclosed by the straight lines
frequencies $\lambda_1,\lambda_2$ and $\lambda_3$, corresponding
to $z'_1, z'_2 ~\textrm{and}~ z'_3$.}

\section{Stable gridding intervals} \label{sec:StableGridding}
This section focuses on the problem how to discriminate
$r_3-$intervals, such that stable polygons may exist therein. The
basic idea presented here tempts to extrapolate this information
from the plot of equation (\ref{eq:singFreqs}), see
Fig.~\ref{fig:kpwexample}. Basically, one would like to link
somehow the stability of the characteristic polynomial
(\ref{eq:basicEquation}) with the number of singular frequencies
on $\Gamma$. This approach is motivated by the fact that an
emerging pair of singular frequencies produces new polygons and
vice-versa.

The following lemma is essential for the main result of this
section. Without loss of generality, the $\Gamma$-region is
assumed to be a Schur-circle. The generalizations for
$\Gamma-$circles are straightforward. Its proof is, however,
beyond teh scope of this article (proof hints: (a) $N$ is always
even, (b) the Mikhailov curve is symmetric to the real axis $\tau$
and (c) the poles on $\Gamma$ are avoided by infinitely small
circles, whereby the Mikhailov plot experiences a phase shift of
$\pm\pi$ at infinity.)

{\it
\begin{lem}\label{lem:MikhailovRealZeros}
Consider the Mikhailov plot of a function $F(e^{j\alpha})$ on the
Schur-circle $\Gamma_1$ and let $F(z)$ possess $J$ poles $\neq \pm
1$ on $\Gamma_1$, a pole of the order $J_+$ at $z=+1$, and a pole
of the order $J_-$ at $z=-1$. If the phase change of the Mikhailov
vector $F(e^{j\alpha})$ over $\Gamma-1$ is $N\pi$ as $\alpha$
changes from $0$ to $+\pi$ without touching them, then it cuts the
real axis at least $Z-$times, where

\begin{equation}
    \label{eq:lemmakPCondition}
Z = \frac{N-J-2-E(J_+)-E(J_)}{2}.\\
\end{equation}

%
%
\end{lem}
}

{\it
\begin{thm}\label{thm:r1GrddingIntervals}
Consider the characteristic polynomial (\ref{eq:basicEquation})
and the Schur-circle $\Gamma_1$. Let\\
\begin{tabular}{ll}
$N$: & order of the polynomial (\ref{eq:basicEquation})\\
$R$: & number of zeros of $A(z)E_\Gamma(z)$ lying inside $\Gamma$\\
$J$: & number of zeros $\neq\pm 1$ of $A(z)E_\Gamma(z)$ lying on $\Gamma$\\
$J_+$: & order of the zero $+1$ of $A(z)E_\Gamma(z)$ \\
$J_-$: & order of the zero $-1$ of $A(z)E_\Gamma(z)$ \\
$Z$: & number of singular frequencies in the interval $0 < \alpha < +\pi$.\\
\end{tabular}

A necessary condition for stability of (\ref{eq:basicEquation}) is
\begin{equation}
    \label{eq:kPCondition}
    Z \geq N-R-\frac{J+E(J_+)+E(J_)+2}{2}.
\end{equation}
\end{thm}
}
 \vspace{0.5cm}

\emph{Proof.~} As shown in Section~\ref{sec:basicDefinitions}, the
equation
\begin{equation}
F(z)=\frac{p(z)}{A(z)E_\Gamma(z)}=0
\end{equation}
decouples parameters $r_1,r_2$ and $r_3$ into the two equations
(\ref{eq:singFreqsLines}) and (\ref{eq:singFreqs}), whereby the
imaginary part represents the generator of singular frequencies.
The Mikhailov plot of function $F(z),
~z=\tau(\alpha)+j\eta(\alpha)), ~-\pi\leq\alpha\leq\pi$ intersects
the real axis exactly at singular frequencies $z'=z(\alpha')$. If
$p(z)$ is $\Gamma-$stable, then according to the principle of
argument on $\Gamma_1$ yields
\begin{equation}
    \label{eq:principleArgumetn}
    \Delta \Phi_F=(N-R)2\pi,
\end{equation}
where $\Delta \Phi_F$ represents the phase change of the function
$F(z)$ on $\Gamma_1$ as $\alpha$ changes from $-\pi$ to $+\pi$,
including $\pm\pi$. Notice that $J+J_++J_-$ zeros of
$A(z)E_\Gamma(z)$ lying on $\Gamma$ are avoided by exclusion via
infinitely small semicircles, i.e. they are considered to be
outside $\Gamma_1$. Thus according to
Lema~\ref{lem:MikhailovRealZeros}, the equation
(\ref{eq:kPCondition}) results if $N$ is substituted by $2(N-R)$
in (\ref{eq:lemmakPCondition}). \hfill\QED

Using this theorem one can directly read from the plot of equation
(\ref{eq:singFreqs}) the $r_3-$interval(s) where stable polygons
may exist. However, the bounds defined by this theorem may be
conservative if stable polygons disappear due to the intersection
of at least three singular lines at one point in the parameter
space ($r_1,r_2,r_3$). The reader is referred to
\cite{bajc:auto:2004}, where a thorough discussion on this item is
provided.

\subsection{Example: (cont)}

{~Consider Schur-stability for $A$ and $B$ given in
(\ref{eq:Azexmaple}) and (\ref{eq:Bzexmaple}). It can be checked
that $A(z)$ possesses three zeros inside the Schur-circle, one
zero at $z=-1$ and one zero outside the Schur-circle. Thus if the
decoupling function $E_\Gamma(z)=z$, see
(\ref{eq:decouplingFunctionCircle}), is used, it follows that
\begin{displaymath}
N=8,~~R=3+1=4, ~~J=1,~~J_+=0,~~\textrm{and}~~J_-=1.
\end{displaymath}
Hence, for stability, $Z\geq 3$ singular frequencies are required
in the interval $0<\alpha<+\pi$. In order to discriminate stable
$r_3$ intervals one should check the plot of the generator of
singular frequencies shown in Fig.~\ref{fig:kpwexample}. The
stable interval is depicted by the grayed strip in
Fig.~\ref{fig:kpwexample}
\begin{displaymath}
-0.52236 < r_3 <0.00290.
\end{displaymath}
Notice that the zoomed plot in Fig.~\ref{fig:kpwexample}
recommends that for $0< r_3 <0.00290$, four additional singular
frequencies appear.

On the other side if the decoupling function $E_\Gamma(z)=1+z^2$
is used, then
\begin{displaymath}
N=8,~~R=3,~~J=3,~~J_+=0,~~\textrm{and}~~J_-=1
\end{displaymath}
i.e. again for stability $Z\geq 3$ singular frequencies are
required in the interval $0<\alpha<+\pi$.

\subsection{Example: PID control}
Now consider PID control of the same plant (\ref{eq:GsDigit}) with
the control law (\ref{eq:PID}). It can be shown that in this case,
\begin{eqnarray}\label{eq:A1B1}
A&=&{z}^{3}+ 10.98\,{z}^{2}+10.98\,z+ 1\\
B&=&0.1 z^6-0.5 z^5+z^4-z^3+0.5 z^2-0.1 z.
\end{eqnarray}

By using $E_\Gamma(z)=z$, it is easily checked that
\begin{displaymath}
N=6,~~R=1+1=2, ~~J=0,~~J_+=0,~~\textrm{and}~~J_-=1
\end{displaymath}
since $A(z)$ has one zero inside the Schur-circle, one zero at
$z=-1$, and the third one outside. Hence, $Z\geq 3$ singular
frequencies within $0<\alpha<+\pi$ are required. However, one can
check that the maximal number of singular frequencies within
$0<\alpha<+\pi$ is $2$, so no PID controller can stabilize the
plant (\ref{eq:GsDigit}).

\begin{figure}[h]
\center {\includegraphics[scale=0.5 ]{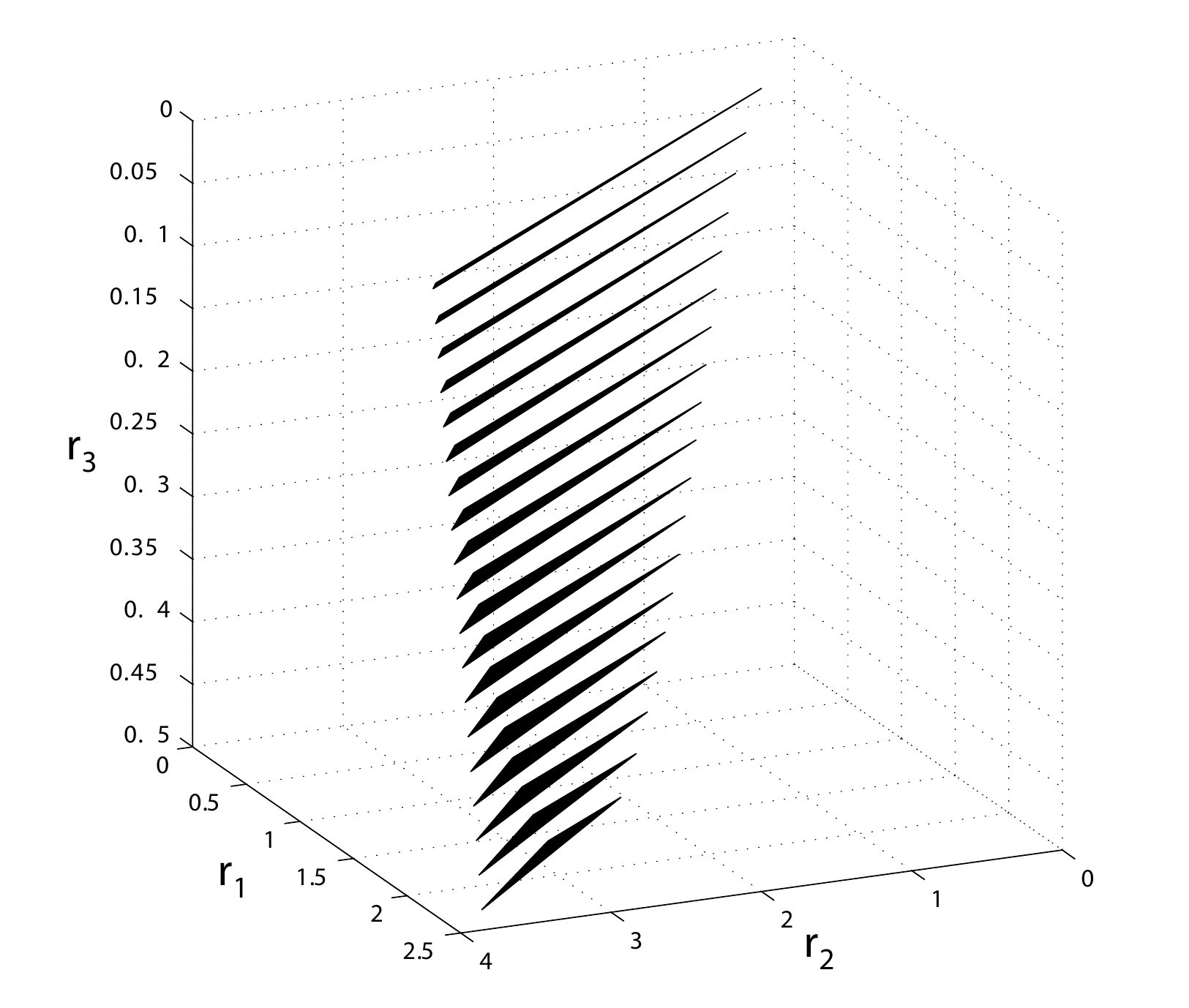}}
\caption{The set of all Schur-stable polygons}\label{fig:ack3}
\end{figure}

\subsection{Example: (cont)}
Having identified the stable intervals, griding of parameter $r_3$
yields stable slices, which compose the $3-$D stability region.
E.g. Fig.~\ref{fig:ack3} shows the whole region in
$(r_1,r_2,r_3)-$ parameter space of three-term controllers,
(\ref{eq:threeTerm}), which stabilize the plant
(\ref{eq:GsDigit}).

\section{Conclusion}
The problem of finding the set of all PID and three-term
stabilizers for a linear discrete-time system is treated in this
paper. The basic result is the transfer and generalization of the
counterpart theory of continuous-time PID stabilizers to
time-discrete domain. The design method is based on the fact that
controller parameters in a linearly transformed parameter space
appear decoupled at singular frequencies. Thereby the design
problem decouples into detection of stable polygons and assertion
of intervals where such polygons may exist. A new simple and
powerful rule is introduced to discriminate such intervals. The
design method for simultaneous stabilization of several operating
points becomes feasible by intersecting convex polygons.

\end{document}